\documentstyle[emulateapj,graphicx]{article}
\def\lcdm  {\rm $\Lambda$CDM\ }
\def\ldcdm  {\rm $\Lambda$DCDM }

\def\msun{{\rm\,M_\odot}}

\def\vol#1  {{{#1}{\rm,}\ }}

\def\etal{et al.\ }

\newcount\refno
\newcount\rfno
\def\eq{$^{\the\refno\ }$\advance\refno by 1}
\def\ad{\advance\rfno by 1}

\def\clock{\count0=\time \divide\count0 by 60
     \count1=\count0 \multiply\count1 by -60 \advance\count1 by \time
     \number\count0:\ifnum\count1<10{0\number\count1}\else\number\count1\fi}

\def\Gcm2{\rm G~cm^2}

\def\beq{\begin{equation}}
\def\eeq{\end{equation}}

\def \date         {\ifcase\month \message{zero} \or
                    January \or February \or March \or April \or May \or June 
                    \or July \or 
                    August \or September \or October \or November \or 
                    December \fi
                    \space\number\day, \number\year}

\begin{document}
\title{Why Are There Dwarf Spheroidal Galaxies?}
\author{Renyue Cen\altaffilmark{1}}
\altaffiltext{1} {Princeton University Observatory, Princeton University, Princeton, NJ 08544; cen@astro.princeton.edu}
\accepted{ }

\begin{abstract}

\end{abstract}

There appears to exist a conflict between the standard structure
formation theory and the existence of the faint
dwarf spheroidal galaxies in the Local Group.
Theoretical considerations suggest that 
a cold dark matter universe 
has been a very hostile place for 
the formation of small galaxies.
In particular, 
gas would not have
been able to cool and condense to form stars
inside dark matter halos with a velocity dispersion $\le 10~$km/s
since before the cosmological reionization epoch.
In other words, one should not expect to see any major star formation
activities in dwarf galaxies with a velocity dispersion $\le 10~$km/s
during the past $12~$Gyr, according to the current structure formation theory.
The real universe, on the other hand, shows that
all dwarf spheroidal galaxies in the Local Group
have velocity dispersions $\le 10~$km/s and 
extended and recent star formation activities are quite common in them. 

This apparent conflict between cold dark matter models and local observations
can be resolved,
if one allows the cold dark matter particles to decay relatively recently.
The resolution comes about
in that the dwarf spheroidal galaxies 
with a velocity dispersion of $\sim 10~$km/s seen today
were born, in this picture, 
in a small fraction 
of halos with a velocity dispersion of $\sim 20~$km/s, where 
gas can be retained and is able to cool and contract to form stars.
The presently observed lower velocity dispersion ($\sim 10~$km/s)
of these dwarf spheroidal galaxies 
is a consequence of the decay of one half of the dark matter in 
and subsequent expansion of the halos since redshift $z\sim 2$.


\section{Introduction}

In the standard picture of structure formation,
over the redshift interval $z\sim 10-100$ a sufficient amount of 
molecular hydrogen may be able to form and enable gas to cool and condense
at temperature $T\ll 10^4$K,
capable of assembling stellar systems with a velocity
dispersion $\ll 10~$km/s (Haiman, Thoul, \& Loeb 1996;
Tegmark \etal 1997).
However, a very small fraction of total baryons collapsed in these
first generation stellar systems would have formed enough stars
to generate enough 
sub-13.6eV photons (to which the universe is largely transparent even then)
to completely photo-dissociate hydrogen molecules
well before enough hydrogen ionizing photons ($>13,6~$eV)
have been produced to reionize the universe.

Subsequently, gas cooling and hence collapsing is not possible in halos
with a virial temperature
$T_{v}<10^4$K (or baryonic masses $<10^8[(1+z_v)/10]^{-3/2}$; 
Haiman, Rees, \& Loeb 1997).
However, an early soft X-ray background
could enhance the formation of 
hydrogen molecules (by providing extra free electrons)
thus may somewhat prolong the period of 
gas cooling in halos with a virial temperature $T_{v}<10^4$K
(Haiman, Abel, \& Rees 2000).
Once hydrogen molecules are destroyed,
star formation becomes inactive 
(there may be exceptional regions enriched with metals where
metal cooling at $T<10^4$ is possible thus some star formation activities
may take place there),
until the nonlinear scale of the density perturbations
reaches a virial temperature $T\sim 10^4$K.

At that point a sufficient number of galaxies in
halos of virial temperature $T\ge 10^4$K 
(with hydrogen collisional excitation cooling being the primary coolant)
quickly form enough stars 
and the universe gets reionized at a very rapid pace 
(Gnedin \& Ostriker 1997).
Between the epoch of molecular hydrogen elimination
and the completion of cosmological reionization
some gas may passively sit in halos with $T_v\le 10^4$K 
but is incapable of cooling and condensing to form stars.
Unfortunately, even this already settled gas will be driven
out of the shallow halo potential wells
(on a timescale of sound crossing time $\ll 1~$Gyr),
when the universe is reionized,
leaving the halos virtually gasless (Barkana \& Loeb 1999)
and depriving them of chance for
any further star formation.

After reionization, the temperature of the intergalactic medium
of the universe is raised to $\ge 10^4$K (Miralda-Escud\'e \& Rees 1994)
and subsequent 
accretion of gas onto halos with $T_v\le 10^4$K is prohibited
(Efstathiou 1992;
Thoul \& Weinberg 1996;  
Quinn, Katz, \& Efstathiou 1996; 
Kepner, Babul, \& Spergel 1997;
Navarro \& Steinmetz 1997; 
Kitayama \& Ikeuchi 2000;
Gnedin 2000b). 
Reionization of once-ionized helium at lower redshift
(perhaps at $z\sim 3$) may still increase the temperature
of the intergalactic medium (Hui \& Gnedin 1997; Schaye \etal 2000),
while additional shock heating due to breaking of long density waves
progressively raises the temperature of the intergalactic medium further 
at somewhat lower redshift (Cen \& Ostriker 1999).

In summary, it does not seem to be possible 
to (bottom-up) form isolated stellar systems with a velocity
dispersion $<10~$km/s since redshift at least as high as $z=6$,
at which the universe seems to have already been reionized (Fan \etal 2000).
The age of the universe then is $\le 1~$Gyr.
In other words, star formation in dwarf galaxies with a velocity
dispersion $\le 10~$km/s
can only occur $\ge 12~$Gyr ago (assuming that the present age
of the universe is $13~$Gyr) and
no later episodes of star formation are allowed. 
This is a generic feature of cold dark matter models
as well as warm or collisional dark matter models.

However, observations of our local neighborhood, the Local Group,
show that there exist a dozen or so dwarf spheroidal
galaxies, all of which have a velocity dispersion $\le 10~$km/s
($8.5\pm 2.0~$km/s seems to bracket the velocity dispersions
of all the Local Group dwarf spheroidal galaxies; van den Bergh 2000).
These dwarf spheroidal galaxies
have complicated star formation histories but it is
clear that extended and recent star formation 
activities are common. 
For example, the star formation in Fornax may be as recent 
as $<1~$Gyr ago (Saviane, Held, \& Bertelli 2000),
so are Sculptor (Monkiewicz \etal 1999) and Sagittarius (Bellazzini, Ferraro,
\& Buonanno 1999);
Carina experienced three significant episodes of star formation at 
$\sim 15$, $\sim 7$ and $\sim 3~$Gyr ago,
with the main burst of star formation 
during the episode $7~$Gyr ago (Hurley-Keller, Mateo, \& Nemec 1998);
In Leo I 70-80\% of stars formed between 7 and $1~$Gyr ago (Gallart \etal 1999).

How can one reconcile this clear theoretical prediction
with solid local observations?
One possible solution is that these dwarf spheroidal galaxies are
produced during the formation of or 
interactions (such as close encounters or mergers) between giant galaxies.
However, this explanation seems unlikely
because the central dark matter density of the observed dwarf 
spheroidal galaxies
appears to be higher than that of giants;
fragments of larger galaxies are unlikely to have a higher density
than their parents (Kormendy \& Freeman 1998).

In this {\it Letter},
we suggest that this apparent conflict 
may be resolved in the decaying cold dark matter model ($\Lambda$DCDM)
recently proposed (Cen 2000).
This is not because gas can be accreted onto
halos with a velocity dispersion $<10~$km/s in $\Lambda$DCDM
rather it is because halos seen today with a velocity dispersion $10~$km/s 
have a velocity dispersion $20~$km/s at birth
in $\Lambda$DCDM, onto which gas accretion is not prohibited 
or already
settled gas would not have been driven out of the halos
in the cosmological reionization phase.

\section{Formation of Dwarf Spheroidal Galaxies in Decaying Cold Dark Model}

Dwarf spheroidal galaxies have very high mass-to-light ratios 
(van den Bergh 2000, Fig 15.3; Hirashita, Tamura, \& Takeuchi 2000)
and thus are dark matter dominated (Faber \& Lin 1983;
Lin \& Faber 1983).
Observations indicate that the stellar mass function in them
is very similar to globular clusters in giant galaxies
down to very small stellar mass, implying that it is unlikely
that the dark matter in the dwarf
spheroidal galaxies may be stellar,
since globular clusters do not contain a significant
amount of dark matter (Feltzing, Gilmore, \& Wyse 1999).
This then brings out our assumption
that dwarf spheroidal galaxies
are cold dark matter dominated stellar systems 
on all scales in the context of cold dark matter models.
Therefore, the dynamics of dwarf spheroidal galaxies is 
mostly determined by the cold dark matter within;
dynamically, baryons evolve passively inside dark matter halos
of dwarf spheroidal galaxies.

The decaying cold dark matter model (Cen 2000)
was recently suggested to alleviate the small-scale power excess
problem at $z=0$ in the canonical 
cosmological constant dominated cold dark matter model 
(Ostriker \& Steinhardt 1995),
which has been remarkably successful on larger scales.
In this model one half of the cold dark matter particles
decay into relativistic particles by zero redshift
with a decay time of $1.4t_0$ ($t_0$ is the current age of the universe).
By $z=2$ about $10\%$ of cold dark matter particles 
have decayed.
Hence, in this model, virialized halos are not static
but evolve at low redshift ($z<2$) when 
a substantial amount of dark matter starts to decay.
Because the dark matter decay time scale is much longer 
than the dynamic time of a halo,
a halo primarily experiences an adiabatic expansion 
when the cold dark matter particles decay at low redshift.
Since the baryons do not play a significant dynamic role,
the dynamics of dark matter halos is particularly simple.
Basically, an isolated halo will grow larger in size (radius)
by a factor of approximately two and its total mass 
becomes smaller by a factor of approximately two,
after one half of the dark matter has decayed.
Therefore, the velocity dispersion of such a halo
is approximately reduced by a factor of two
from $z\sim 2$ to $z=0$.

As a consequence, a halo seen today with a velocity dispersion of $10~$km/s
had a velocity dispersion of $20~$km/s at birth and 
had maintained this velocity dispersion until $z\sim 2$ in $\Lambda$DCDM.
Unlike halos with a velocity dispersion of $10~$km/s, where
accretion of photoionized gas is completely suppressed,
gas accretion onto halos with a velocity dispersion of $20~$km/s 
is permitted.
Besides, throughout the cosmological
reionization phase these halos would have retained gas that was settled in
before cosmological reionization.
This may resolve the apparent conflict 
between the existence of extended star
formation activities in dwarf spheroidal galaxies 
and the cold dark matter model.

\section{Discussion}

Although star formation is, in principle,
not prohibited in halos with a velocity dispersion of $20~$km/s
and could occur over an extended period throughout the life of the universe,
gas accretion onto them may be significantly hindered 
by photoionization heating and other heating processes.
In fact, simulations have shown that photoionization
heating can cause suppression of gas accretion onto halos
with a velocity dispersions up to $50-70~$km/s 
(Thoul \& Weinberg 1996; 
Quinn, Katz, \& Efstathiou 1996; 
Navarro \& Steinmetz 1997;
Gnedin 2000b).
But detailed hydrodynamic simulations with sufficient
mass and spatial resolutions to resolve the halos in question
are required to quantify this effect definitively.

However,
even if gas accretion onto halos with a velocity dispersion of $20~$km/s
is completely suppressed after the cosmological reionization epoch,
these halos may still have enough gas that was
accreted at earlier times;
these halos could have accreted gas before
the cosmological reionization epoch and retained the gas,
unlike halos with a velocity dispersion of $\le 10~$km/s 
(Barkana \& Loeb 1999).
It is relevant to discuss the effect of supernovae from starbursts in 
dwarf spheroidal galaxies.
Mac Low \& Ferrara (1999) have shown that mass ejection efficiency
is very low for galaxies with mass $M>10^7\msun$.
Since the dwarf spheroidal galaxies today have mass $M>2\times 10^7\msun$
(and higher at high redshift in our model),
it seems possible that they have retained most of their gas 
over their lifetime.
This would then have permitted extended star formation activities
in them, although
star formation activities could be
intermittent 
due to internal supernova heating (Burkert \& Ruiz-Lapuente 1997;
Hirashita 2000);
this is consistent with the observed star formation histories 
of local dwarf spheroidal galaxies.

It is worth noting that the observationally
inferred velocity dispersion thus mass of the dwarf spheroidal
galaxies is based on the assumption
that these galaxies are in dynamical equilibrium.
If they are not in dynamic equilibrium, then their mass would be
lower and dark matter content
may not be as high, which would exacerbate the problem considered here;
i.e., without dark matter halos, the star formation activities 
of the dwarf spheroidal galaxies would be even more difficult
to understand. 
Considerations of tidal perturbations of dwarf spheroidal galaxies
by giants prefer the existence of a deep potential well 
of dark matter (Hirashita, Kamaya, \& Takeuchi 1999);
in other words, without dark matter halos, it will be very difficult
to maintain the orderly existence of 
dwarf spheroidal galaxies in the vicinities of large galaxies 
as in the Local Group environment.
Under the assumption of dynamical equilibrium 
one finds that the average density within
the virial radius of the dwarf spheroidal galaxies
is approximately $10^6\rho_{crit}$ (where  
$\rho_{crit}$ is the present critical density of the universe),
indicating an early formation epoch ($z\ge 10$) of these halos
(e.g., Peebles 1997).
This early epoch of formation has the virtue
of collecting gas easily before cosmological reionization
and does not necessarily require substantial
accretion of gas at later epochs, as discussed above.
Using Press-Schechter (1974) theory one estimates that
the number density of halos with a velocity dispersion $>20~$km/s
is $10^3-10^4~$Mpc$^{-3}$ at $z=10$ in the $\Lambda$DCDM model,
significantly higher than the number density of the Local Group
dwarf spheroidal galaxies ($\sim 1~$Mpc$^{-3}$).
Clearly, a large number of halos 
with a velocity dispersion $20~$km/s
could have formed before cosmological reionization.

Klypin \etal (1999) have shown that the abundance of halos with a velocity
dispersion of $\ge 20~$km/s in the \lcdm model at $z=0$
is roughly a factor of $3$ higher than the 
abundance of the observed dwarf spheroidal galaxies with a velocity
dispersion of $\sim 10~$km/s in the Local Group.
Assuming that the only effect of introducing 
decaying cold dark matter particles
is to reduce the velocity dispersion of halos by a factor of two,
the difference between their simulations
and observations implies that only about $1/3$ of halos with
a velocity dispersion of $\sim 20~$km/s in the \ldcdm model 
that have not been cannibalized  
end up as dwarf spheroidal galaxies with a velocity dispersion 
of $\sim 10~$km/s as seen in the Local Group.
The fate of the remaining $2/3$ of dwarf spheroidal galaxy halos with a
velocity dispersion of $\ge 10~$km/s has interesting
consequences.
One possibility is that they may have been tidally disturbed
and thus are more dispersed to be detected easily
perhaps at stages later than the Sagittarius dwarf galaxy 
(Ibata, Gilmore, \& Irwin 1994) or is already part of the halo
stars of giants.
The second possibility is that these dwarf spheroidal galaxies
are just further away and they are not yet detected due to 
their low surface brightness.
The third possibility is that they were neither tidally disrupted
nor formed enough stars.
In this case, it is unclear whether they are gas rich or not.
Perhaps likely is that the gas content of these
``dark" dwarf spheroidal galaxies is strongly environment dependent;
those closer to the giants may 
suffer effects such as ram pressure stripping.
It is also possible that these ``dark" dwarf galaxies
may contain mostly ionized gas, thus their gas may not be easily
detectable in HI 21~cm surveys (e.g., Young 2000).

Interestingly, there is an intriguing coincidence.
In the vast majority of all the observed dwarf spheroidal galaxies
the star formation rate declines sharply 
about $10~$Gyr ago (Grebel 1998; Mateo 1998; Gnedin 2000a),
which corresponds to redshift $z\sim 2$
in the \ldcdm model (Cen 2000).
Incidentally, this is about the redshift after which
cold dark matter particles start to decay significantly
and hence dwarf galaxy halos start to expand significantly.
Two effects could arise.
First, the expansion of the halos cause a simultaneous expansion of the 
baryonic gas thus reduces the baryonic surface density.
Second, the lowering of the halo velocity dispersion 
further reduces the accretion 
rate of photoionized gas onto the halos.
Both effects reduce star formation rate
and could have caused the sharp decline of star formation rate
in the observed dwarf spheroidal galaxies $\sim 10~$Gyr ago.

Finally, the scenario presented here may be tested by future observations.
It is predicted that there should be no
dwarf spheroidal galaxies with a velocity dispersion 
of $\le 10~$km/s at $z\ge 2$;
there should be a minimum velocity dispersion
of approximately $20~$km/s for 
dwarf spheroidal galaxies at $z\ge 2$.
The dwarf spheroidal galaxies should be
more compact at $z\ge 2$ than today;
they should be smaller in size by a factor of $\sim 2$
and surface brightness higher by a factor of $\sim 4$ (perhaps more
considering stellar evolution) at $z\ge 2$ than today.
NGST may be able to detect such dwarf spheroidal galaxies
at $z\sim 2$ (Stockman \& Mather 1999).

\section{Conclusions}

A host of theoretical arguments indicate  that
no significant star formation activity should have taken place in 
dwarf spheroidal 
galaxies with a velocity dispersion $\le 10~$km/s in the past $12~$Gyr,
in any cold (or warm or collisional) dark matter universe.
Observations, on the other hand, tell us that
all dwarf spheroidal galaxies in the Local Group
have a velocity dispersion $\le 10~$km/s and 
much more recent (than $12~$Gyr ago)
star formation activities in them are common.
Therefore, we face a conflict between theory and observations.

We suggest that this apparent conflict may be resolved in
the decaying cold dark matter model.
In this model these dwarf spheroidal galaxies were born 
in halos with a velocity dispersion 
of $\sim 20~$km/s, in which gas may be retained and/or
accreted, and cooling and condensation could occur
for an extended period of time.
Thus, extended and recent star formation activities
in them would have been possible.
The presently observed lower velocity dispersion 
($\le 10~$km/s) of these dwarf spheroidal 
galaxies results from the decay of one half of the dark matter in 
and subsequent expansion of the halos since redshift $z\sim 2$.

Future observations such as by NGST may be able to
detect high redshift dwarf spheroidal galaxies.
It is predicted by the current scenario
that they should be smaller, have higher surface brightness
and higher velocity dispersion at high redshift.

\acknowledgments
This research is supported in part by grants AST93-18185
and ASC97-40300.
I thank Zoltan Haiman and Jerry Ostriker for many useful comments
that significantly improved the paper.

\end{document}